\documentclass[modern]{aastex7}

\newcommand\ha{{H$\alpha$}}
\newcommand\hb{{H$\beta$}}

\newcommand\kms{\:\rm{km\,s^{-1}}}

\newcommand\peryr{\rm{yr}^{-1}}
\usepackage[utf8]{inputenc}
\usepackage{natbib}
\usepackage{comment}
\usepackage{todonotes}
\usepackage{graphicx}
\usepackage{lineno}
\usepackage{comment}
\usepackage{amssymb}

\shorttitle{Optical Evolution of Tycho's SNR}
\shortauthors{Winkler, Putko, \& Blair}

\begin{document}

\title{The Optical Evolution of Tycho's Supernova Remnant over Three Decades}

\correspondingauthor{P. Frank Winkler}

\author[0000-0001-6311-277X]{ P. Frank Winkler}
\affil{Department of Physics, Middlebury College, Middlebury, VT 05753, USA}
\email{winkler@middlebury.edu}

\author[0000-0002-1576-0083]{Joseph Putko}
\affil{Department of Physics, Middlebury College, Middlebury, VT 05753, USA}
\affil{Leidos Security Enterprise Solutions, Tewksbury, MA, 01876 USA}
\email{josephputko@gmail.com}
%\and

\author[0000-0003-2379-6518]{William P. Blair}
\affil{The William H. Miller III Department of Physics and Astronomy, Johns Hopkins University, 3400 N. Charles Street, Baltimore, MD 21218, USA}
\email{wblair@jhu.edu}

\begin{abstract}
%==============================================================
We report a series of  images of Tycho's supernova remnant at eight epochs extending over thirty years: 1986-2016.  In addition to our \ha\ images, we have obtained  matched continuum images which we subtract to reveal faint emission, including a far more extensive network of optical knots and filaments than reported previously.  The deepest images also show an extremely faint, fairly diffuse arc of emission surrounding much of the circumference of Tycho to the southeast and south, coinciding with the rim of the radio/X-ray shell.  We have measured proper motions for 46 filaments, including many fainter ones near Tycho's outer rim.  Our measurements are generally consistent with previous ones by \citet{kamper78}, but ours have far greater precision.   Most optical filaments at the shell rim have expansion indices reasonably consistent with the Sedov value (0.40), while the interior filaments have somewhat smaller values, as expected.   From the combination of proper motions of filaments at the shell rim and shock velocity at the same positions, one should be able to calculate the distance to Tycho by simple geometry.   Determination of the shock velocity from  broad Balmer-line profiles is subject to model uncertainties, but the availability of dozens of such filaments with a range of conditions offers the possibility to substantially improve the distance determination for Tycho.

\smallskip
\center{\it{Dedicated to the memory of Janet Winkler, who made it all fun throughout these decades.}}
%==============================================================
\end{abstract}


\section{Introduction\label{intro}}
%==============================================================

\begin{figure}
\epsscale{0.60}
\plotone{ 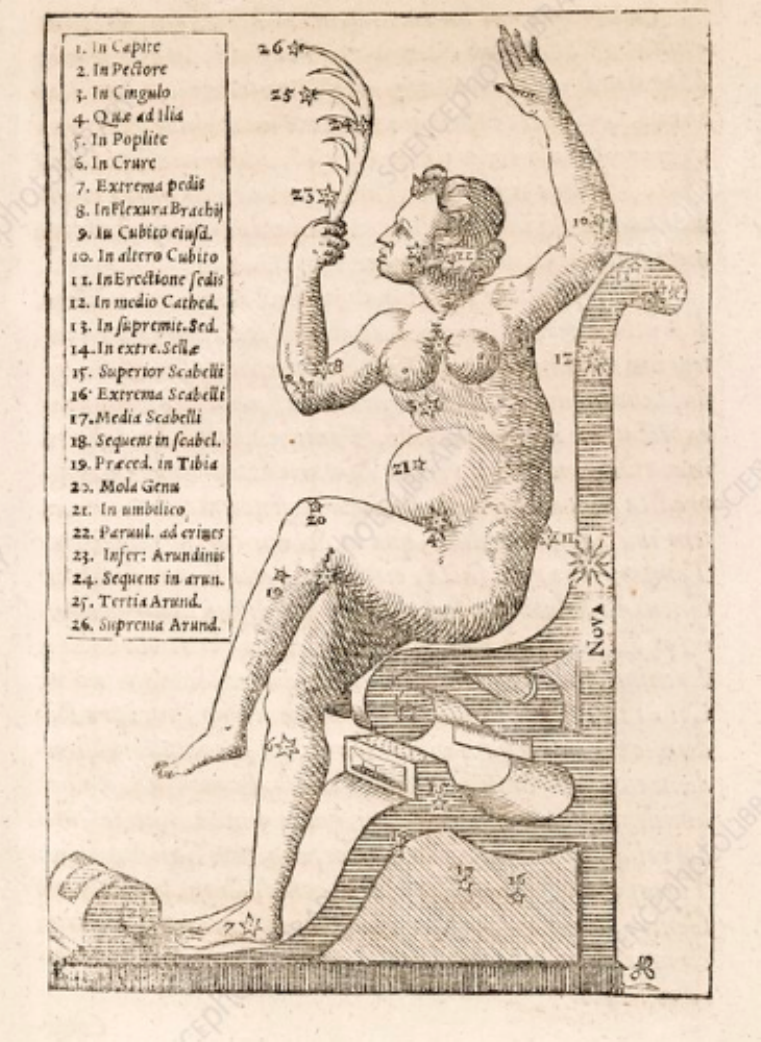}
\caption{The ``new star" of 1572, which appears to be almost blasting Cassiopeia out of her chair (from Tycho's posthumously published {\it Progymnasmata}, 1602).  (This figure does not appear in the {\em Ap J}).\label{woodcut}}
\end{figure}

The supernova of 1572 CE is probably the best documented by contemporary observers of all the historical supernovae of the pre-telescopic era.  In early November 1572 a new star suddenly appeared in Cassiopeia and rapidly brightened to rival Venus  (Fig.\ref{woodcut}).
 It then gradually faded until it became invisible in March 1574.  The most extensive set of observations was carried out by Tycho Brahe, who reported them in his {\it De Stella Nova} \citep{tycho1573} and more completely in his posthumously published {\it Progymnasmata} \citep{tycho1602}, which also incorporated observations by others.  Based on these observations, \citet{baade45} reconstructed the light curve, with peak magnitude $m_V \approx -4.0$, and concluded that the 1572 event had been a supernova, almost surely of Type I. (At the time, supernovae were classified only into types I and II, but more recent classification schemes have clearly identified it as a Type Ia based on its reconstructed light curve.)  SN1572 was conclusively shown to be of Type Ia when \citet{krause08} spectroscopically observed its light echo, at that time recently discovered by \citet{rest08b}.
\citet{ruiz-lapuente23} reviewed the historical observations in a modern context, in which SNIa have become important tools for cosmology. 

The first published report of the remnant of Tycho's SN is a 1407 MHz   observation from the one-mile radio telescope at Cambridge by \citet{baldwin67}.  His map shows a shell with striking circular symmetry,  about 8\arcmin\ in diameter. This source is also known as 3C10, and the SNR more generally is also designated as G120.1+1.4.   According to \citet{van-den-bergh71}, Baade detected optical emission from the remnant on a red-sensitive plate taken from the newly commissioned 5m Hale telescope at Mt. Palomar  in 1949, though there seems to be no actual published record of this image.  Follow-up plates in 1955 and 1957 showed that the filaments of the remnant have large proper motions.  \citet{van-den-bergh71} took another 5m plate, noted that the optical filaments lie near the north and east limbs of the radio shell, speculated that these were thin sheets seen nearly edge-on, and measured the proper motions of several filaments  to be $0\farcs 20 - 0\farcs 24\; {\rm yr}^{-1}$.  A more definitive study of Tycho's proper motions was carried out by \citet[][henceforth KV78]{kamper78}, using a series of seven plates from the Hale 5m telescope.  They measured the motions of 14 features in Tycho, obtaining values ranging from $0\farcs 15\; \peryr$ to $0\farcs 23\; \peryr$.  We will discuss their results further and compare them with our own in Sec.\ \ref{results} of this paper.

Tycho's SNR was among the first sources detected in the early days of X-ray astronomy  by \citet{friedman67}, and it appears in the Second {\em Uhuru} catalog \citep{giacconi72}.  More recent multi-wavelength observations will be discussed in Sec.\ 5.

\begin{figure}
\epsscale{1.15}
%\plottwo{09_cont_subt.pdf}{ultra_faint.pdf}
\plottwo{ 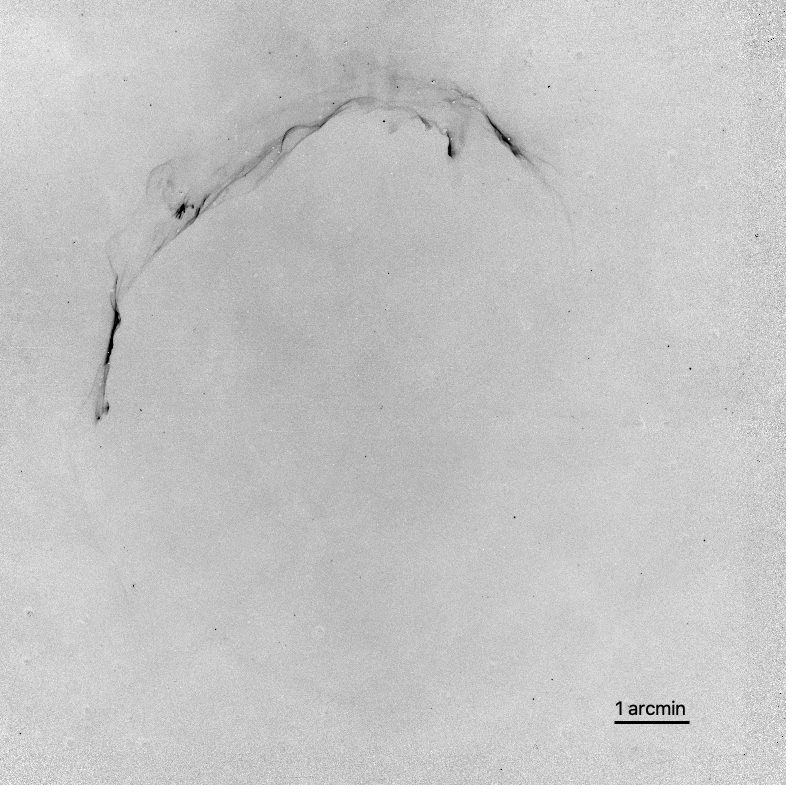}{ 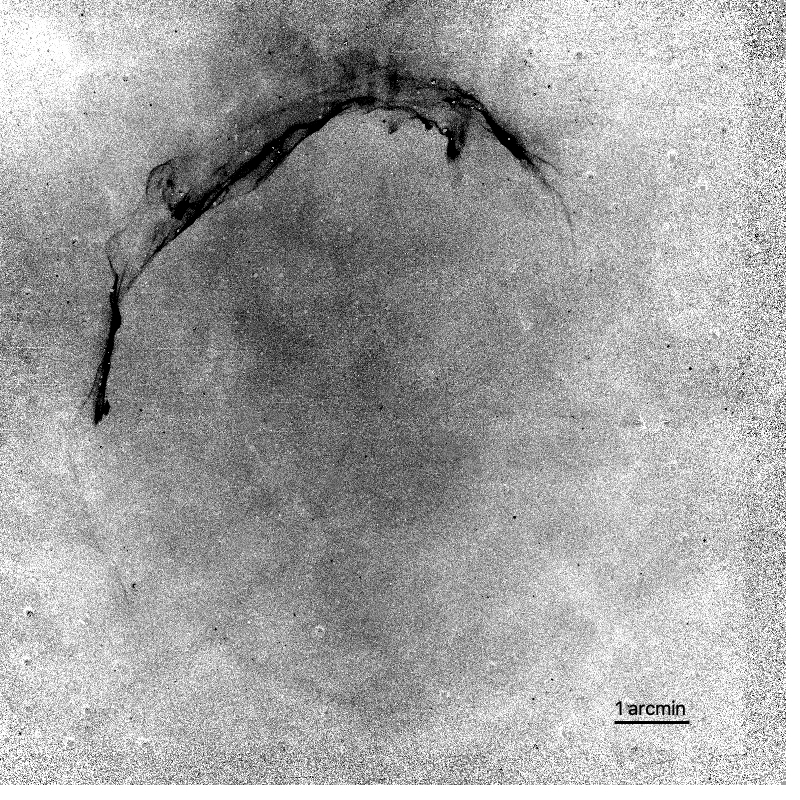}
\caption{{\em Left:} Tycho's SNR in H$\alpha$ at epoch 2009, after continuum subtraction. North is up; east is to the left. {\em Right:} The same image as at left but with a different display scale: now the familiar portion of the optical remnant appears significantly saturated to bring out extremely faint H$\alpha$ emission in the S and SE. %The radial profile cut in the south was used for Figure~\ref{fig-ultra-faint_profile}. 
\label{fig-2009}}
\end{figure}

\citet{minkowski59} reported that the only emission line visible in the Tycho spectrum was \ha.  The first published spectrum was by \citet{kirshner78}, and showed both \ha\ and \hb, but no other lines.  Balmer line emission results when cold neutral atoms ahead of the SN shock drift through the shock, where they encounter a population of hot protons.  The post-shock neutrals can undergo charge exchange with  fast protons, resulting in excited fast neutral atoms.  These then emit broad Balmer lines, and the width of these lines gives a measure of the shock velocity \citep{chevalier78, chevalier80}.  \citet{heng10} has reviewed the physics of this emission in considerable detail.

This paper is organized as follows:  In Sec.\ \ref{obsns} we present the observations that comprise the data set we use for this study, and in Sec.\ \ref{meas} we describe the methods by which we extract one-dimensional profiles across numerous filaments and then measure their shifts from one epoch to another.  Sec.\ \ref{results} gives the results of our study and   compares these  with previous proper motion measurements.  Sec.~\ref{multi} discusses further implications of our study, and places it in the context of multi-wavelength observations of Tycho.  Sec.\ \ref{distance} discusses how our results may be used to improve the still uncertain determination of the distance to Tycho.  Finally, in Sec.\ \ref{summary} we provide a brief summary.

%==============================================================

\section{Observations and Data Reduction \label{obsns}}
%==============================================================
Narrow-band imaging observations of Tycho's SNR in H$\alpha$ were carried out at eight epochs from 1986 to 2016 as detailed in Table ~\ref{tbl-obs}.  All the imaging for our study was carried out using CCDs, thus dating almost from the earliest common use of CCDs for astronomy.  Two of the images used in this study have been previously published: The 1986 image is shown in \citet{kirshner87}, and the 1990 image in \citet{smith91}, in both cases to illustrate the slit positions for spectra reported in those papers.
All images from all epochs were processed using standard IRAF\footnote{IRAF is distributed by the National Optical Astronomy Observatory, which is operated by AURA, Inc., under cooperative agreement with the National Science Foundation.} procedures for bias-subtraction and flat-fielding.

\begin{deluxetable}{lcccccc}
%\rotate
\tabletypesize{\small}

\tablecolumns{9}
\tablewidth{0pt}
\tablecaption{Summary of  Observations\label{tbl-obs}}
\tablehead{
		\colhead{Date}  & 
		\colhead{Telescope}  & 
		\colhead{CCD}  &
		\colhead{Scale}  &  
		%\colhead{$\lambda_{\mathrm{c}}$(\AA)}  & 
		%\colhead{$\Delta\lambda$(\AA)}  & 
		\colhead{Exposure }  &
		\colhead{Seeing }   & 
		\colhead{Observers}\\
		%\colhead ($^{\prime\prime} \mathrm{pixel}^{-1}$)} & \colhead (s) & ($^{\prime\prime}$)} }
		\colhead{} &
		\colhead{} &
		\colhead{} &
		 \colhead {($^{\prime\prime}\; \mathrm{pixel}^{-1}$)} & 
	%	\colhead {($foo$)} & 
		 \colhead {(s)} & 
		 \colhead {($^{\prime\prime}$)} & 
		 \colhead{} 
		 }

\startdata
1986 Sep 15\tablenotemark{a}  & Mayall 4.0 m   & TI2  & 0.29    & 2 $\times$ 600   & 1.1 & Blair \\
1988 Nov 10\tablenotemark{a,b}   & Mayall 4.0 m   & TI2   & 0.29   &  3 $\times$ 960   & 1.0 &  Fesen\tablenotemark{c} \\
1990 Sep 28   & Mayall 4.0 m   & T2KB   & 0.46    & 1800 + 1560 & 1.4 & Blair \\
1996 Nov 17   & Mayall 4.0 m   & T2KB   & 0.46     & 4 $\times$ 600   & 1.5   & Winkler, Williams\tablenotemark{c}, Eriksen\tablenotemark{c} \\
2003 Aug 30   & MDM 2.4 m   & SITe 2K   & 0.28      & 3 $\times$ 2500 & 0.8 & Fesen\tablenotemark{c} \\
2007 Oct 04   & KPNO 2.1 m   & T2KB   & 0.30     & 6 $\times$ 900   & 1.1 & Winkler, Vaughan\tablenotemark{c} \\
2009 Sep 18--19\tablenotemark{d}   & WIYN 3.5 m   & OPTIC   & 0.14   &    20 $\times$ 600 & 0.7 & Winkler, Blair \\
%\\  &   &   &   & 6852   & ?   & 20 $\times$ 200   &   & 
2016 Oct 23\tablenotemark{d,e}--  &Gemini-N 8.2m  & GMOS (e2V)  & 0.146 &  48 $\times$  500 & 0.7 - 1.2 &  Winkler (PI)\\
\quad2017 Jan 29 
		\enddata
\tablenotetext{a}{Only the northeast portion of the remnant was observed.}
\tablenotetext{b}{A three-panel mosaic.}
\tablenotetext{c}{Robert A. Fesen, Dartmouth College; Benjamin F. Williams, now Univ. Washington; Kristoffer Eriksen now Los Alamos National Laboratory; Matthew Vaughan.}
%\tablenotetext{d}{Mosaic data.}
\tablenotetext{d}{A narrow-band continuum image was also obtained.}
\tablenotetext{e}{Mosaic of 4 quadrants, each in 2 perpendicular orientations.  Program GN-2016B-Q-82, queue observations.}

\end{deluxetable}

%\end{document}

The eight epochs were all precisely aligned using 172 reference stars from the UCAC4 catalog \citep{zacharias13}. Stars with negligible proper motions were selected, as this permitted use of this same list for all epochs.\footnote{Much of this work was completed before the far more extensive GAIA catalog became available; however, comparison of results using the two catalogs for the 2016 epoch shows negligible differences.}  We calculated a world coordinate system (WCS) for each frame of the 2009 epoch using the IRAF task \textit{msctpeak} 
%(in the MSCFINDER subpackage of MSCRED) 
to find the centroids of the reference stars and to fit them to a common tangent-plane projection. We chose the 2009 epoch as the reference image since it has excellent seeing and  minimal distortion. 
%After copying this WCS to the other epochs, 
Similarly, we calculated a WCS for each frame within each of the other  epochs, also via \textit{msctpeak}, allowing for non-linear terms as required to achieve excellent fits with no systematic residuals.   \added{Comparison of several stars at multiple epochs shows that the images are aligned to far better than single-pixel accuracy.}
%The projection and fitting geometry parameters within \textit{msctpeak} can be set for linear transformations or transformations of higher order, the latter to account for distortion. For each epoch, the parameters that gave fit residuals appearing more randomly distributed and smaller residuals before deleting any reference stars (to further improve the rms dispersion) were used.

%Having all the epochs on a common image coordinate system facilitates measuring proper motions.  
We arbitrarily defined a standard coordinate system: a linear tangent-plane projection centered at R.A.~(2000.) $=$ 00\textsuperscript{h}25\textsuperscript{m}20.00\textsuperscript{s}, Decl.~(2000.) $= +64^{\circ}08'30.00''$, with a scale of $0.200^{''} \mathrm{pixel}^{-1}$. Each individual frame was transformed to the standard one using IRAF, where the fitting geometry parameters were set to match those used previously in \textit{msctpeak} for each epoch. At this point all the frames from each epoch were combined to obtain a single-epoch image for each of the eight epochs.  

In addition, narrow-band red continuum images were  taken in 2007, 2009, and 2016, to be used for star subtraction. Star subtraction is useful to clearly identify where filaments are contaminated by stars, and it permits faint H$\alpha$ emission to be seen more clearly. The red continuum images were PSF-matched with the H$\alpha$ images at the same epoch, scaled, and then subtracted fom the \ha\ images to remove most of the stars.   For each of the earlier epochs, where no continuum images were obtained, the 2009 continuum image was PSF-matched, scaled, and used for star subtraction.  

The 2009 epoch, our deepest  image taken in a single field rather than mosaicked, is shown (after star subtraction) in Fig.~\ref{fig-2009}.    For the first time, emission surrounding almost two-thirds of the entire rim can be seen. The right panel of Fig.~\ref{fig-2009} shows this image with the familiar optical emission significantly saturated to bring out extremely faint optical emission, particularly in the southeast quadrant of the shell. This previously undetected emission agrees well with that seen in radio and X-ray images, including the ``blow-out"  \citep{reynoso97, katsuda10}.   Also seen in Fig.~\ref{fig-2009} is very faint, diffuse emission near the center of the shell and just outside to the north.  We see comparable emission in both the 2007 and 2016 images, so it must be real, rather than an artifact.  Similar emission, though somewhat more structured, can be seen for the SN\thinspace Ia remnant SN1006 \citep{winkler03}.  A shock precursor could be responsible for the emission just outside the primary shell.

%\todo[inline]{Consider this?:  ``Deep optical images may reveal faint optical counterparts to the radio signatures of the R-T instability taking place in Tycho" (Velazquez 1998)}

\begin{figure}
\plotone{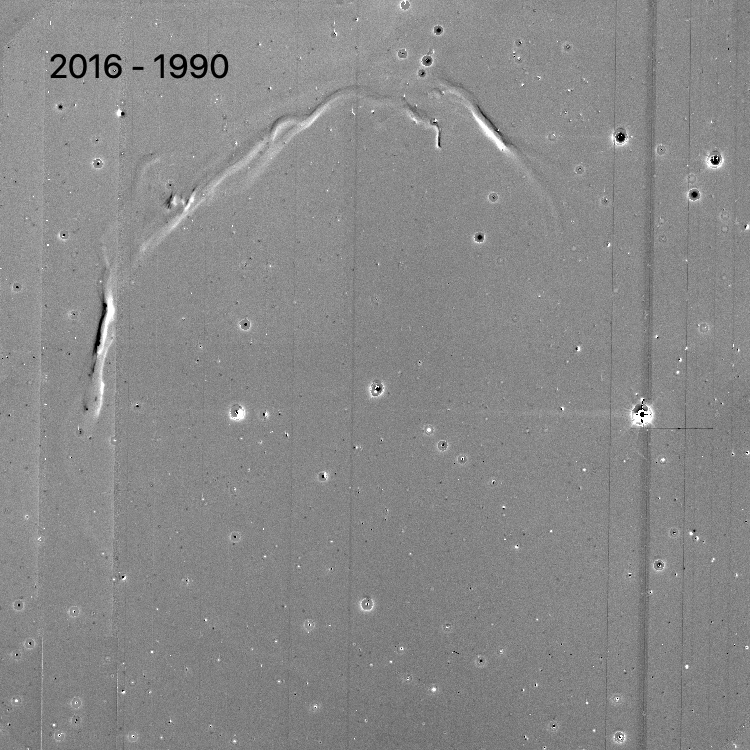} \caption{Difference image between those from epochs 2016 and 1990.  At the later epoch the brightest filaments appear as black, while at the earlier one they are white.  The expansion should be evident. \label{diff}}
\end{figure}

\begin{figure}
\includegraphics[angle=-0,width=15.5cm]{ 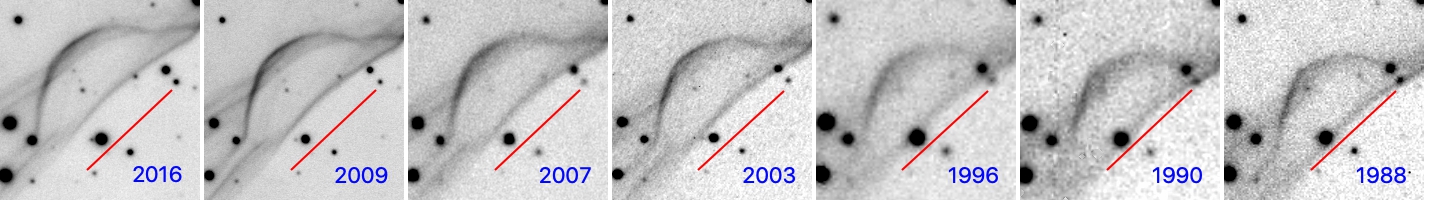}  %Had to correct figure size and orientation (I think because I had unusual page setup settings when it was made).
\caption{A sequence of images  of coherent H$\alpha$ filaments  Az19.1 and  Az20.0 in the northeast quadrant of Tycho's SNR. The bar  is in the identical position in each frame.  Each frame is 20\arcmin\ square. The sequence proceeds right-to-left. \label{fig:NE_coherent_movie}}
\end{figure}

\begin{figure}
\includegraphics[angle=-0,width=15.5cm]{ 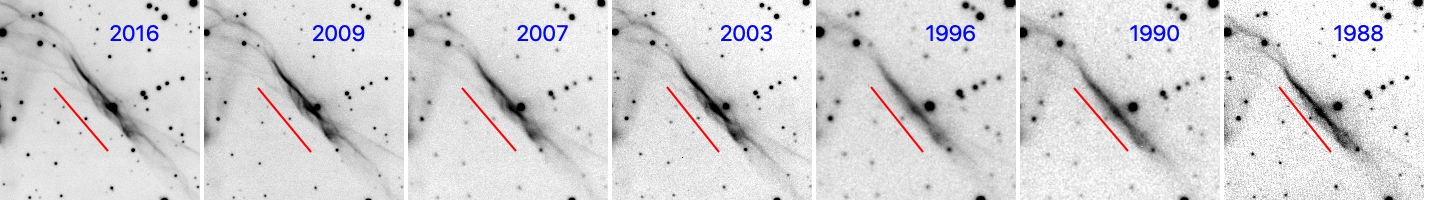}
%\plotone{NW_movie.pdf}
\caption{A sequence of images (proceeding right-to-left)  of the brightest H$\alpha$ filament in the northwest quadrant of Tycho's SNR, \added {comprising filaments Az334.4, Az336.9, and Az339.4.}  The red bar  is in the same position in each frame. Each frame is 40\arcmin\ square.  \label{fig-NW_movie}}
\end{figure}

As a clear demonstration of Tycho's expansion, we have taken the difference between the (continuum-subtracted) images from 2016 and 1990, the two epochs showing the full remnant that are separated by the longest baseline, shown in Fig.~\ref{diff}.
To highlight all the observations used in this study and the clear proper motions of Tycho's optical filaments over our long time baseline, we show two ``movie images" in Figs.~\ref{fig:NE_coherent_movie} %(box 1 in Fig.~\ref{fig-2009}),
and ~\ref{fig-NW_movie}. %(box 2 in Fig.~\ref{fig-2009}), and~\ref{fig-NW_movie} (box 3 in Fig.~\ref{fig-2009}). 
We have chosen the \ha\ images before continuum subtraction, so  the stars can serve as fiducial markers to make the filaments' motions more obvious.  In each movie image, the same region  is shown for all epochs. %Figure~\ref{fig-E_rim} contains the same field with and without stars, highlighting successful star subtraction. 
%Even though image quality is not consistent, from Fig;~\ref{E_rim} especially it is obvious that significant changes in filament structure and relative brightness have occurred. While this complicates measuring the proper motions, such evolution is itself of interest, as we discuss in Section 5.  
Incidentally, these image sequences also serve to trace the development of CCD image technology over these decades.

%==============================================================

\section{Proper Motion Measurements \label{meas}}

Some of the brightest filaments do not move coherently; instead, they may evolve in morphology or appear to ``leap" ahead between epochs.  %Figs.\ \ref{E_rim}, especially, and \ref{N_rim} show several filaments of this sort.  
Since it would be difficult or impossible to obtain unambiguous measurements for these evolving filaments, we concentrated instead on ones that \textit{do} appear to move coherently throughout our thirty-year time span, and especially on ones that are reasonably linear in morphology.  These are typically fairly faint and oriented roughly tangentially to the SNR shell.  

%In Fig.s\ 4 and 5 we show two example images and projection regions; in Fig.\ 5 we show the profiles from both regions, shifted to give the best match, and in Fig.\ 3 we show a plot of the shifts at all epochs and the linear fits.

For each of these linear filaments, we used DS9 to define a ``projection" region, oriented orthogonal to the filament and typically $6\arcsec$  wide and $20\arcsec$ - $40 \arcsec$ long (enough to easily encompass all the epochs). Identical regions were defined for all epochs.  On the continuum-subtracted images, we integrated along the filament (to improve the signal-to noise) to obtain a 1-D profile as a function of the distance across it for each filament at each epoch.     \added{We then measured the optimum shifts between epochs by minimizing the chi-squared sum of the shifted differences between profiles at different epochs, using the Igor Pro software\footnote{see https://www.wavemetrics.com/} for plotting and curve-fitting.}  We adopted the 2009 epoch as our reference, since it was one of the two deepest, and was taken with a single (dithered) field, hence had quite a uniform background.  (The somewhat deeper 2016 image was a mosaic of four overlapping fields, which was obvious in the background with a hard stretch.)  Finally, we plotted the shifts as a function of epoch for each filament, and obtained a linear fit to measure the proper motion.

 %   In Figs.~\ref{fig:NE_coherent_movie} and \ref{fig-NW_movie} we show a sequence of images for two small regions, readily illustrating filament motions.  
 
\begin{figure}
\plotone{ 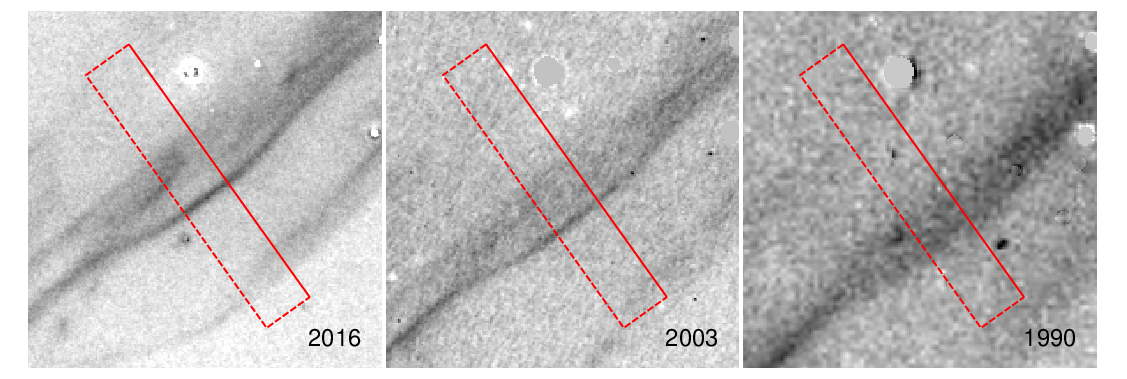}
\caption{An example of the  roughly linear filament Az31.6 and the region defined to obtain its profile, at three different epochs.  Fig.\ \ref{measured_shifts} shows profiles of this filament at all 8 epochs.\label{n11_triple}}
\end{figure}

\begin{figure}
\epsscale{1.16}
\plottwo{ 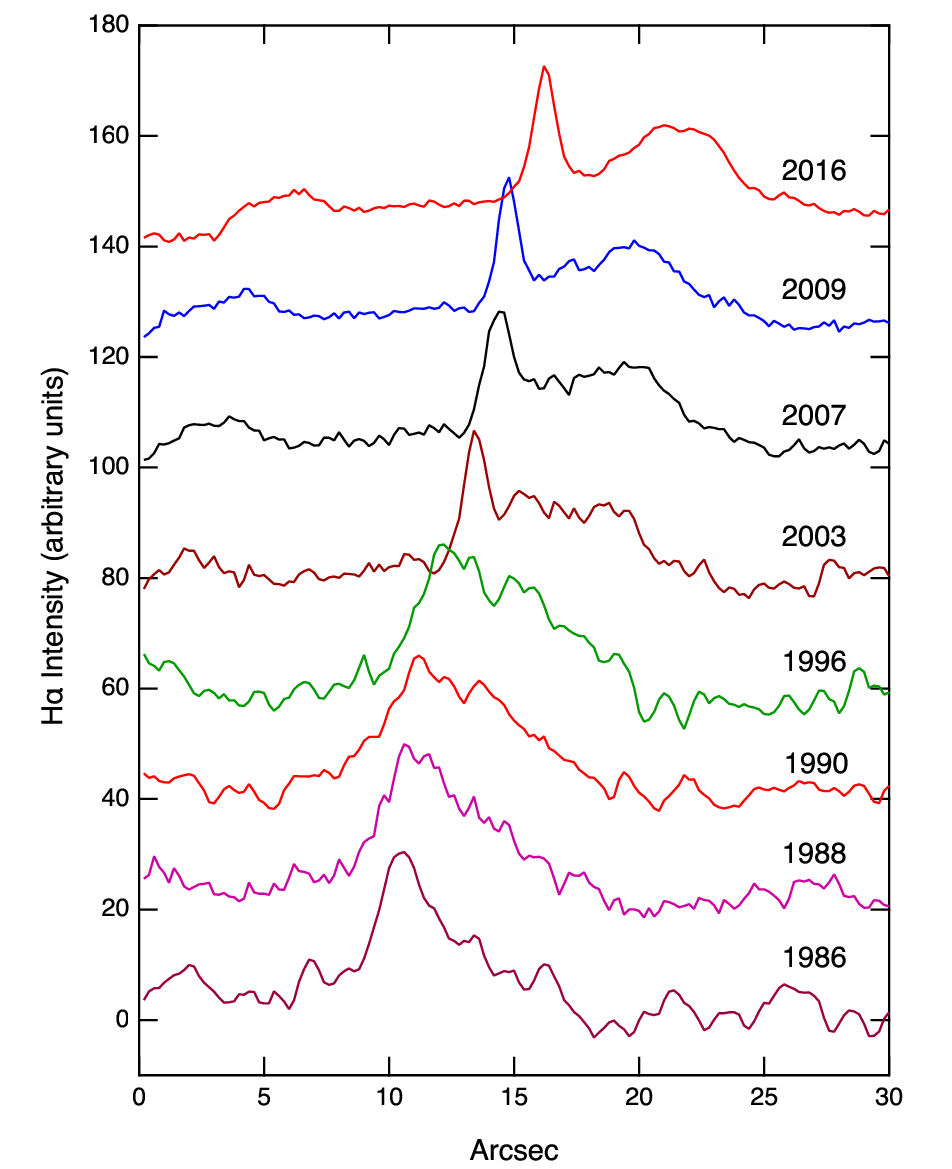}{ 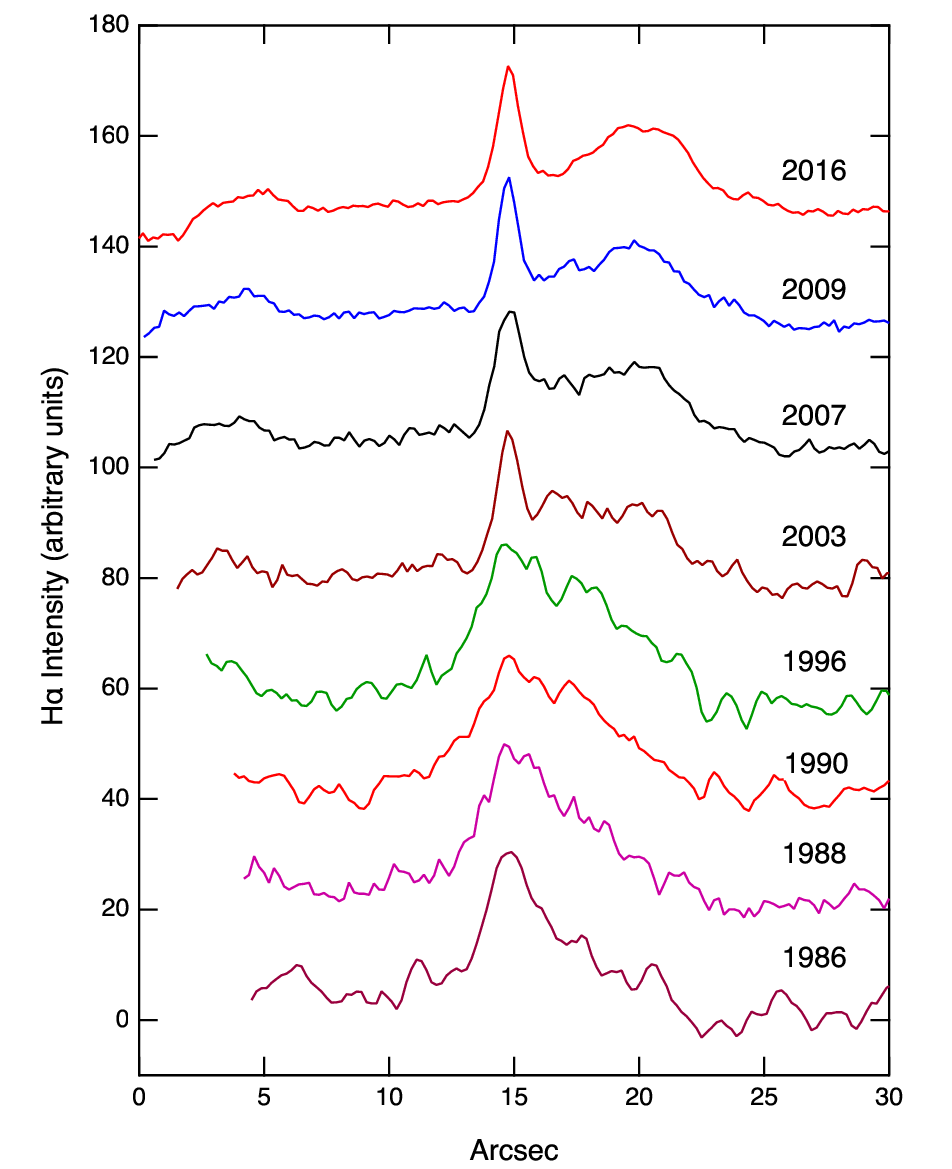}
\caption{{\em Left:}  A sequence of profiles of the well-defined filament Az31.6 in the NE of the Tycho SNR at all 8 epochs.  {\em Right:} The same sequence, now shifted to compensate for proper motions, such that the filament location at all epochs coincides with that in 2009. %The radial profile cut in the south was used for Figure~\ref{fig-ultra-faint_profile}. 
\label{measured_shifts}}
\end{figure}

\begin{figure}
\epsscale{0.7}
\plotone{ 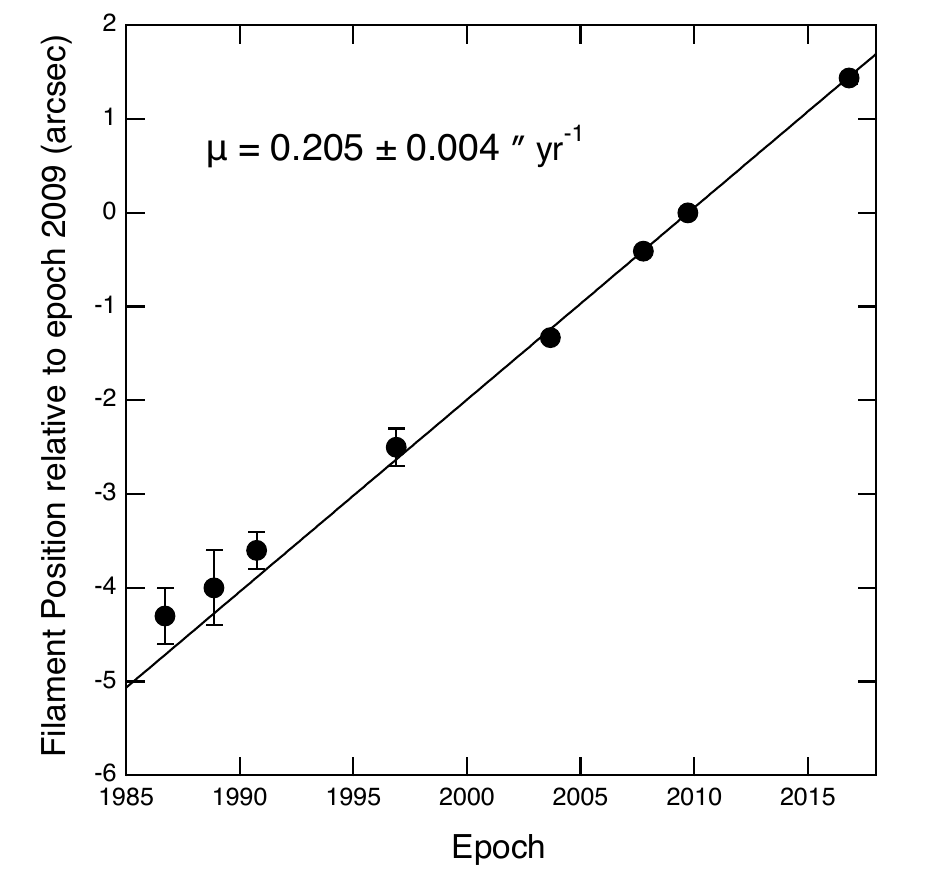}
\caption{Linear fit to the measured shifts for the same filament shown in Figs.\ref{n11_triple} and \ref{measured_shifts}. \label{pm_fit}}
%\label{n11_fit}}
\end{figure}

Fig.~\ref{n11_triple} shows  another example (a fainter filament at three  epochs), with the   region used to extract its profiles. Fig.~\ref{measured_shifts}, left, shows the sequence of eight profiles of the filament in Fig.\ \ref{n11_triple} and on the right the same profiles, shifted to compensate for proper motions to match the 2009 epoch. Finally,  Fig.\ \ref{pm_fit}  shows a plot of the shifts for this filament at all epochs and the linear fit to give its proper motion.

%==============================================================

%\subsection{Measurement of shifts for irregular knots/filaments}

Fifty-two filament regions were visually identified for proper motion measurement.  We  avoided ones where bright stars that were incompletely subtracted precluded obtaining clean profiles, as well as most of those that showed obvious evolutionary changes (though we did measure some bright filaments for comparison with KV78, see Sec. 4.1).   
%Figures~\ref{fig-NE_coherent_movie} and~\ref{fig-NW_movie} are examples of coherent filaments, well-suited for measurement.

%\todo[inline]{We have measured filaments always up to epoch 2009, often leaving out earlier epochs.  What about leaving out later epochs for filaments that seem coherent at earlier epochs?}

\section{Results \label{results}}
\subsection{Proper Motion Measurements}
Table~\ref{tbl-results} summarizes our proper motion measurements for the Tycho filaments, most of which  did not show evolutionary changes.\footnote{Table~\ref{tbl-results} also includes six filaments which {\em did} show acceleration/deceleration and/or evolutionary changes over our 30-year baseline.  All these are noted in the table and discussed in Sec. 4.2.}
We designated these by their azimuth, measured counter-clockwise from north.  For each, the table gives the distance (in arcsec) from our adopted expansion center  of R.A.\ (2000.) $= $ 00\textsuperscript{h}25\textsuperscript{m}19.90\textsuperscript{s}, decl.\ (2000.)$\,= +64^{\circ}08'18\farcs 20$. 
 \citet{ruiz-lapuente04b} determined this to be the center of the near-circular X-ray shell based on {\em Chandra} images, and it has been used by \citet{williams16} for their radio/X-ray proper motion study.\footnote{An alternative center, also based on {\em Chandra} images,  was measured by \citet{warren05}, who minimized the ellipticity of the shell.  This was used by {\citet{katsuda10} in their measurement of X-ray expansion.  The Warren-Katsuda center is about 5\arcsec\ southwest of the Ruiz-LaPuente-Williams one.  \added{More recently, \citet{millard22} have estimated the kinematic center based on the expansion of X-ray knots throughout  Tycho.} We have chosen the Ruiz-LaPuente-Williams value so we can directly compare  our optical measurements with the radio-X-ray ones of \citet{williams16}.}}
%\citet{warren05} determined this center  by minimizing the ellipticity of the Tycho shell as observed in X-rays by {\em Chandra}, which is the closer to circular and  sharper than that observed in radio or optical images.\footnote{\citet{katsuda10} used the same center for their X-ray expansion study of Tycho.  This  is about 5\arcsec\ southwest of the expansion center adopted by \citet{williams16}, who used the center determined from {\em Chandra} images  by \citet{ruiz-lapuente04} through an unspecified method.}
The position of each filament is given at epoch 2009 (our reference epoch), and the motion was measured as described in Sec.\ \ref{meas}.  

\added{For the uncertainty in the proper motion, we must also consider uncertainty associated with the possible mis-alignment among the images at different epochs in addition to the statistical uncertainty associated with the fits. Though small, this can still be significant.  By purposely introducing small image shifts and doing comparisons, we estimate that the images are aligned to, at worst, 0.1 pixel = $0\farcs02$. Since misalignment errors should be random among the different epochs, a fit involving $n_{epochs}$ should introduce a net uncertainty of $0\farcs 02/\sqrt{n_{epochs}}$.  We have combined this in quadrature with the statistical uncertainty from the fitting process to obtain the overall uncertainty in the proper motions, as given in column 5 ($\mu $) of Table ~\ref{tbl-results}.}

The expansion index, $m$, is defined as the current expansion rate divided by the lifetime mean expansion: the current radius over  age of the remnant at the time of the 2009 observations = 436.8 years. 
A Sedov remnant would have $m=0.4$, while completely free expansion would of course have $m = 1$.  The filaments we have measured show motions of $0\farcs 157\;\peryr$ to $0\farcs 327\;\peryr$, and $m$ from 0.310 to 0.607.\footnote{We have excluded one highly anomalous filament, Az47.3.  This is an extremely bright filament at our later epochs, and is probably a region where the SN shock has recently encountered high-density ISM.}

\startlongtable
\begin{deluxetable}{ccccccc}
\tabletypesize{\small}
% \rotate
\tablecolumns{7}
\tablewidth{0pt}
\tablecaption{Expansion Measurements of Filaments with Constant Proper Motion\label{tbl-results}}
\tablehead{
		\colhead{Azimuth\tablenotemark{a}}   & 
		%\colhead{PFW}  & 
		\colhead{Radius\tablenotemark{b}}   & 
		\colhead{R.A. (2000.)}    &
		\colhead{Decl. (2000.)}    &
		\colhead{\phn\phn$\mu$\tablenotemark{c}}    &
		%\colhead{Sigma$_\mu$\tablenotemark{c}}    &
		\colhead{$m$\tablenotemark{d}}    &
		%\colhead{Sigma$_m$\tablenotemark{d}}    &
		\colhead{Epochs} 
		}
\startdata
3.2 & 242.0 & 0:25:22.00 & +64:12:19.9 & 0.179  $ \pm $    0.020 & 0.323  $  \pm  $ 0.034 & 7\\
6.1 & 258.1 & 0:25:24.14 & +64:12:34.8 & 0.202  $ \pm $    0.027 & 0.342  $  \pm  $ 0.042 & 5\\
8.9 & 248.6 & 0:25:25.77 & +64:12:23.9 & 0.266  $ \pm $    0.012 & 0.467 $ \pm  $ 0.012 & 4\\
10.6 & 233.1 & 0:25:26.45 & +64:12:07.4 & 0.165  $ \pm $    0.015 & 0.310 $ \pm  $ 0.023 & 6\\
13.3 & 225.5 & 0:25:27.84 & +64:11:57.8 & 0.208  $ \pm $    0.007 & 0.403 $ \pm  $ 0.003 & 8\\
17.6 & 246.5 & 0:25:31.33 & +64:12:13.2 & 0.265 $ \pm $ 0.011 & 0.469 $ \pm  $ 0.014 & 6\\
19.1 & 217.0 & 0:25:30.79 & +64:11:43.3 & 0.216 $ \pm  $ 0.007 & 0.435 $ \pm  $ 0.001 & 8\\
20.0 & 229.3 & 0:25:31.91 & +64:11:53.6 & 0.233 $ \pm  $ 0.009 & 0.443 $ \pm  $ 0.009 & 8\\
27.0 & 203.3 & 0:25:34.06 & +64:11:19.5 & 0.237 $ \pm  $ 0.008 & 0.509 $ \pm  $ 0.008 & 8\\
31.6 & 212.6 & 0:25:36.96 & +64:11:19.4 & 0.205 $ \pm  $ 0.008 & 0.421 $ \pm  $ 0.008 & 8\\
35.0 & 213.6 & 0:25:38.65 & +64:11:13.2 & 0.227 $ \pm  $ 0.009 & 0.465 $ \pm  $ 0.010 & 8\\
38.6 & 212.6 & 0:25:40.19 & +64:11:04.3 & 0.242 $ \pm  $ 0.008 & 0.496 $ \pm  $ 0.005 & 8\\
42.3 & 216.4 & 0:25:42.21 & +64:10:58.1 & $e$ &  &  \\
45.8 & 262.3 & 0:25:48.73 & +64:11:20.7 & 0.261 $ \pm  $ 0.008 & 0.435 $ \pm  $ 0.005 & 8\\
47.3 & 223.8 & 0:25:45.09 & +64:10:49.8 & 0.081 $ \pm  $ 0.008 & 0.158 $ \pm  $ 0.005 & 8\\
53.1 & 214.0 & 0:25:46.11 & +64:10:26.6 & 0.287 $ \pm  $ 0.008 & 0.586 $ \pm  $ 0.007 & 8\\
56.2 & 218.6 & 0:25:47.72 & +64:10:19.7 & $e$ &  &  8\\
58.4 & 253.1 & 0:25:52.89 & +64:10:30.8 & 0.253 $ \pm  $ 0.012 & 0.437 $ \pm  $ 0.018 & 8\\
59.2 & 220.3 & 0:25:48.88 & +64:10:11.2 & $e$ &  &  8\\
63.3 & 234.1 & 0:25:51.92 & +64:10:03.1 & 0.21 $ \pm  $ 0.013 & 0.391 $ \pm  $ 0.016 & 5\\
63.9 & 246.0 & 0:25:53.71 & +64:10:06.1 & 0.209 $ \pm  $ 0.010 & 0.371 $ \pm  $ 0.009 & 5\\
66.8 & 236.0 & 0:25:53.09 & +64:09:50.8 & 0.252 $ \pm  $ 0.009 & 0.467 $ \pm  $ 0.006 & 6\\
69.9 & 231.7 & 0:25:53.18 & +64:09:37.5 & 0.246 $ \pm  $ 0.008 & 0.465 $ \pm  $ 0.005 & 7\\
74.6 & 225.3 & 0:25:53.12 & +64:09:18.0 & 0.157 $ \pm  $ 0.013 & 0.304 $ \pm  $ 0.021 & 8\tablenotemark{g} \\
77.5 & 225.5 & 0:25:53.57 & +64:09:06.8 & 0.204 $ \pm  $ 0.011 & 0.395 $ \pm  $ 0.016 & 8\tablenotemark{g} \\
79.9 & 225.3 & 0:25:53.82 & +64:08:57.3 &$f$ &  &  \\
85.4 & 225.7 & 0:25:54.30 & +64:08:36.2 &$e$ &  &  \\
90.9 & 235.5 & 0:25:55.39 & +64:08:14.3 & $f$ &  &  7\\
104.3 & 256.1 & 0:25:57.81 & +64:07:14.4 & 0.258 $ \pm  $ 0.017 & 0.439 $ \pm  $ 0.024 & 5\\
110.8 & 256.3 & 0:25:56.45 & +64:06:46.9 & 0.267 $ \pm  $ 0.025 & 0.454 $ \pm  $ 0.039 & 4\\
117.4 & 259.0 & 0:25:55.02 & +64:06:18.7 & 0.281 $ \pm  $ 0.015 & 0.473 $ \pm  $ 0.020 & 6\\
127.3 & 262.2 & 0:25:51.72 & +64:05:39.2 & 0.271 $ \pm  $ 0.017 & 0.451 $ \pm  $ 0.024 & 4\\
137.2 & 236.0 & 0:25:44.34 & +64:05:24.9 & 0.231 $ \pm  $ 0.032 & 0.428 $ \pm  $ 0.056 & 4\\
148.7 & 235.8 & 0:25:38.61 & +64:04:56.6 & 0.256 $ \pm  $ 0.019 & 0.475 $ \pm  $ 0.031 & 5\\
163.8 & 235.3 & 0:25:29.94 & +64:04:32.3 & 0.327 $ \pm  $ 0.032 & 0.607 $ \pm  $ 0.057 & 5\\
171.4 & 242.9 & 0:25:25.43 & +64:04:18.0 & 0.306 $ \pm  $ 0.027 & 0.551 $ \pm  $ 0.045 & 4\\
312.6 & 188.3 & 0:24:58.69 & +64:10:25.6 & 0.174 $ \pm  $ 0.015 & 0.403 $ \pm  $ 0.026 & 4\\
313.9 & 203.2 & 0:24:57.51 & +64:10:39.1 & 0.203 $ \pm  $ 0.022 & 0.436 $ \pm  $ 0.043 & 4\\
317.6 & 200.6 & 0:24:59.20 & +64:10:46.4 & 0.158 $ \pm  $ 0.013 & 0.344 $ \pm  $ 0.019 & 5\\
322.0 & 227.9 & 0:24:58.39 & +64:11:17.8 & 0.181 $ \pm  $ 0.015 & 0.346 $ \pm  $ 0.023 & 5\\
326.3 & 228.5 & 0:25:00.47 & +64:11:28.3 & 0.250 $ \pm  $ 0.009 & 0.478 $ \pm  $ 0.006 & 5\\
329.8 & 222.8 & 0:25:02.74 & +64:11:30.7 & 0.211 $ \pm  $ 0.008 & 0.414 $ \pm  $ 0.007 & 7\\
332.6 & 221.4 & 0:25:04.32 & +64:11:34.8 & 0.206 $ \pm  $ 0.008 & 0.407 $ \pm  $ 0.005 & 7\\
334.4 & 226.6 & 0:25:04.89 & +64:11:42.4 & 0.218 $ \pm  $ 0.008 & 0.420 $ \pm  $ 0.003 & 7\\
336.9 & 231.6 & 0:25:06.00 & +64:11:51.1 & 0.221 $ \pm  $ 0.008 & 0.417 $ \pm  $ 0.005 & 7\\
339.4 & 236.2 & 0:25:07.16 & +64:11:59.3 & 0.234 $ \pm  $ 0.008 & 0.432 $ \pm  $ 0.003 & 7\\
340.4 & 230.6 & 0:25:08.03 & +64:11:55.4 & 0.215 $ \pm  $ 0.010 & 0.406 $ \pm  $ 0.007 & 5\\
343.9 & 249.8 & 0:25:09.25 & +64:12:19.1 & 0.242 $ \pm  $ 0.008 & 0.423 $ \pm  $ 0.005 & 7\\
347.0 & 252.8 & 0:25:11.15 & +64:12:24.5 & 0.196 $ \pm  $ 0.009 & 0.338 $ \pm  $ 0.009 & 7\\
353.4 & 227.9 & 0:25:15.91 & +64:12:04.7 & 0.192 $ \pm  $ 0.010 & 0.369 $ \pm  $ 0.010 & 6\\
356.4 & 233.3 & 0:25:17.68 & +64:12:11.1 & 0.228 $ \pm  $ 0.011 & 0.427 $ \pm  $ 0.012 & 6\\
359.0 & 266.6 & 0:25:19.21 & +64:12:44.8 & 0.288 $ \pm  $ 0.019 & 0.473 $ \pm  $ 0.029 & 5\\
\enddata

\tablenotetext{a}{Designated by azimuth, measured eastward from north.}
\tablenotetext{b}{Distance (in arcsec) from center at 0:25:19.90, +64:08:18.20, at epoch 2009.}
\tablenotetext{c}{Measured proper motion, in $\arcsec\,\peryr$.}
\tablenotetext{d}{Expansion index, $m \equiv \mu/(R/t)$, where  $t=$ age at epoch 2009 = 436.8 yr.}
%\tablenotetext{e}{Profile measurable at fewer than 4 epochs; no $\mu$ reported.}
\tablenotetext{e}{Filament has evolved noticeably.}
\tablenotetext{f}{Filament has decelerated noticeably.}
\tablenotetext{g}{Proper motion measured for epochs 1990-2003 only; see Sec.~\ref{distance}.}

\end{deluxetable}

\begin{figure}
\epsscale{1.3}
\plotone{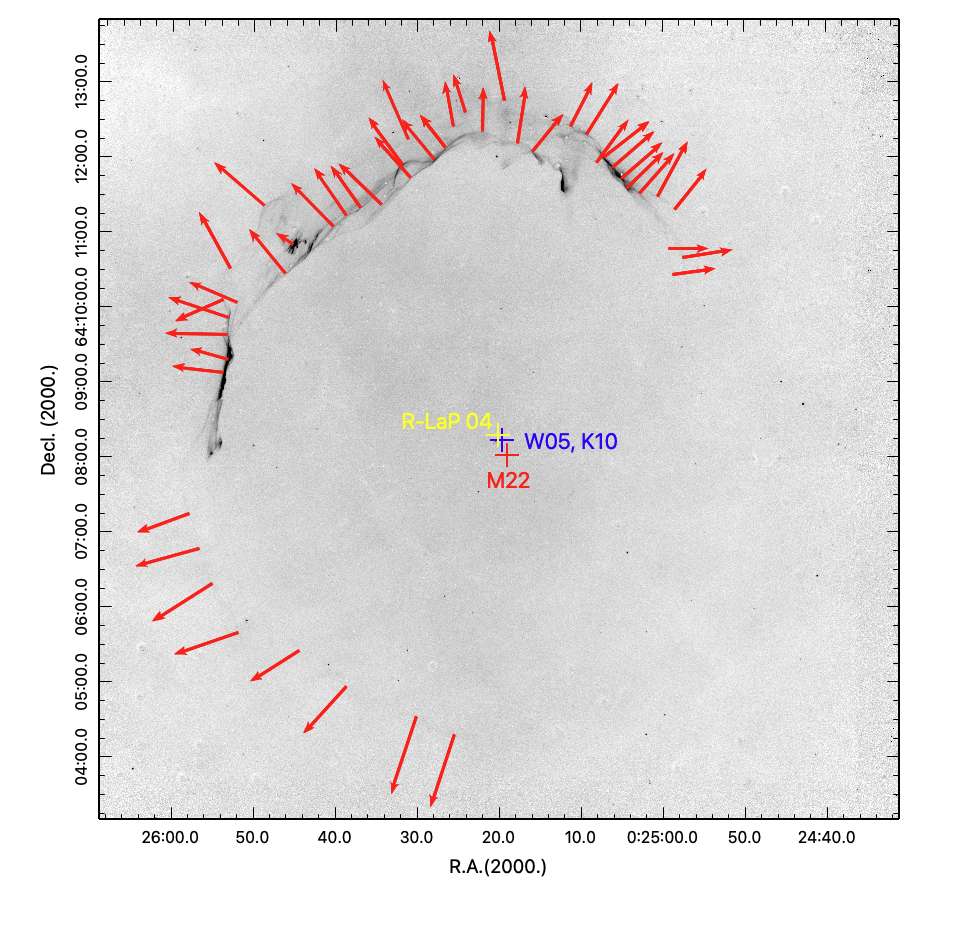}
\caption{The continuum-subtracted 2009 image, with vectors indicating proper motions extrapolated 200 years into the future at the present rate. We do not show vectors for much of the bright eastern complex (azimuth $\sim 80\degr - 90\degr$, see Fig.\ \ref{E_rim}) plus a few other filaments  where the filaments have decelerated or their morphology has evolved too  much for reliable measurement.  The yellow cross indicates the position of our adopted center \citep{ruiz-lapuente04, williams16}. \added{Also shown are center estimates by \citet{warren05} and \cite{katsuda10}, in blue, and by \citet{millard22}, in red.}
\label{extrapolation}}
\end{figure}

Fig.~\ref{extrapolation} shows the measured proper motions projected into the future by 200 years.  
All are generally outward, of course, but there is far from perfect radial symmetry, just as the filament geometry departs substantially from circular.  The fastest filaments are the extremely faint ones along the southeastern rim.   This comes as no surprise:  optical filaments form where the SN shock encounters neutral interstellar gas, and denser material generally means brighter emission.  Thus these faintest filaments are probably regions of low density.  And if the density is generally low in these directions, the shock will have slowed less than in higher density regions.

%\include{table_results.tex}

%\todo[inline]{Should decelerating or evolving filaments be omitted from table 2?  Or should we give pm values from the last 4 epochs? }

\subsection{Comparison with Previous Proper Motion Measurements}

\begin{figure}
\epsscale{1.15}
\plotone{ 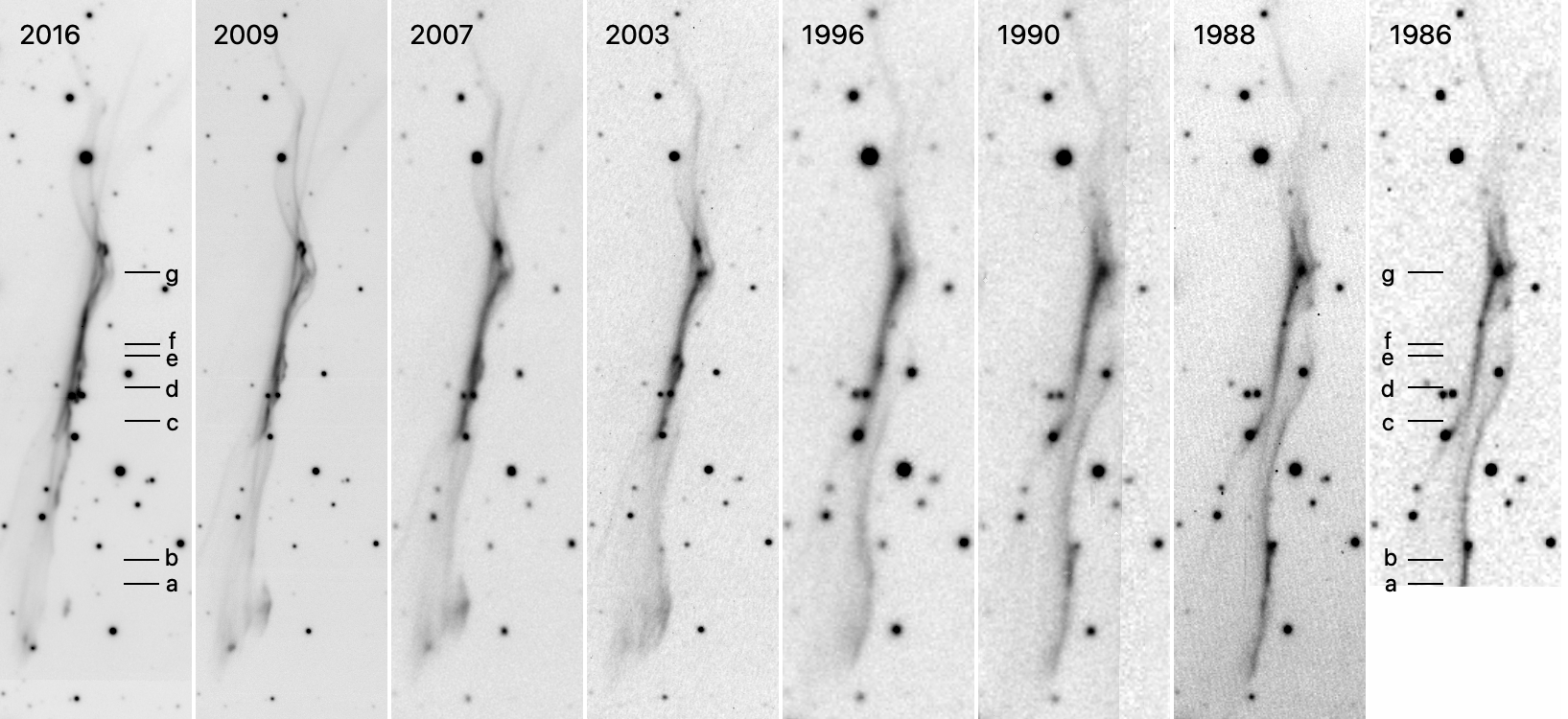}
%\plottwo{E_bright_movie.pdf}{E_faint_movie.pdf}
\caption{A sequence of images of the eastern rim of Tycho's optical SNR (progressing from right to left, same as the direction the filaments move) made from H$\alpha$ images at eight epochs over 30 years.  %Markers A, B, and C (each 15$^{''}$ long) are in the same position in each frame and are used in the text as reference points. The field of view in the 1986 epoch captured less of the remnant. Motion is largely to the east and slightly north (up). 
Both the generally eastward (outward) motion of the filaments and their evolution over time are apparent. The letters a-h indicate features where KV78 measured proper motions.  One may also note the improvement in image quality, resulting from the improvement in instrumentation over this 30-year period. 
The field of each individual image is $40\arcsec \times 150\arcsec$, and the orientation is N up, E left.
\label{E_rim}}
\end{figure}

The definitive previous    study of Tycho's optical proper motions was done by KV78, who used photographic plates from the Palomar 5m at seven epochs spanning 1949 to 1977. They did measurements at 14 positions (which they labeled $a$ through $n$)  along the eastern and northern rims of the radio shell.  In Fig.~\ref{E_rim} we show the brightest filaments in Tycho, a complex network along the eastern rim, comprising KV78 features a - g (their Fig.~2). In our far deeper CCD images, several of their knots or filaments clearly do not move coherently, but instead may evolve in morphology or appear to jump suddenly from one epoch to the next.  
 For example, KV78's knot g, the brightest of the features they measured, and still the brightest in 1986, almost completely disperses into fainter individual filaments by 2016.  On the other hand, the central region comprising KV78 features  c-f brightens substantially from 1986 to 2016.  And the southern extreme of this complex, outside our field in 1986 but seen as relatively bright in our 1988 image, completely dissipates into several far fainter filaments by 2016.

%Figs.\ \ref{E_rim}, especially, and \ref{N_rim} show several filaments of this sort.    This is probably an indication that the SN shock leaves one dense patch of ISM behind and moves rapidly ahead until it encounters another dense patch.

\begin{figure}
\epsscale{1.}
\plottwo{ 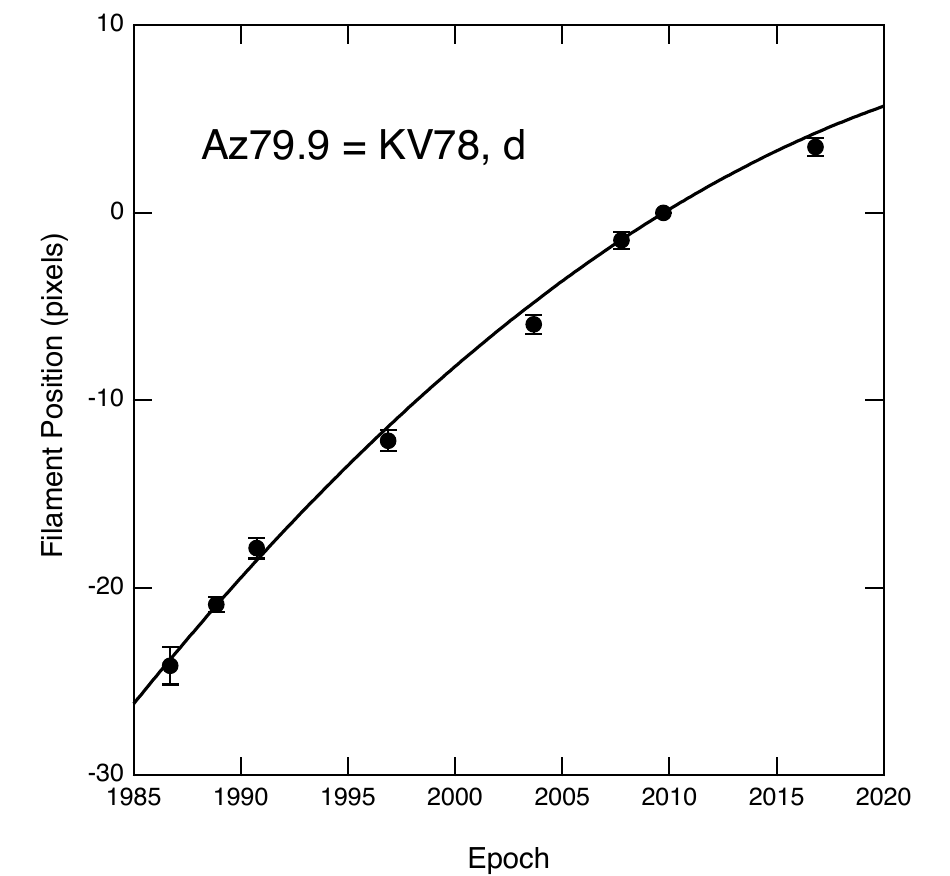}{ 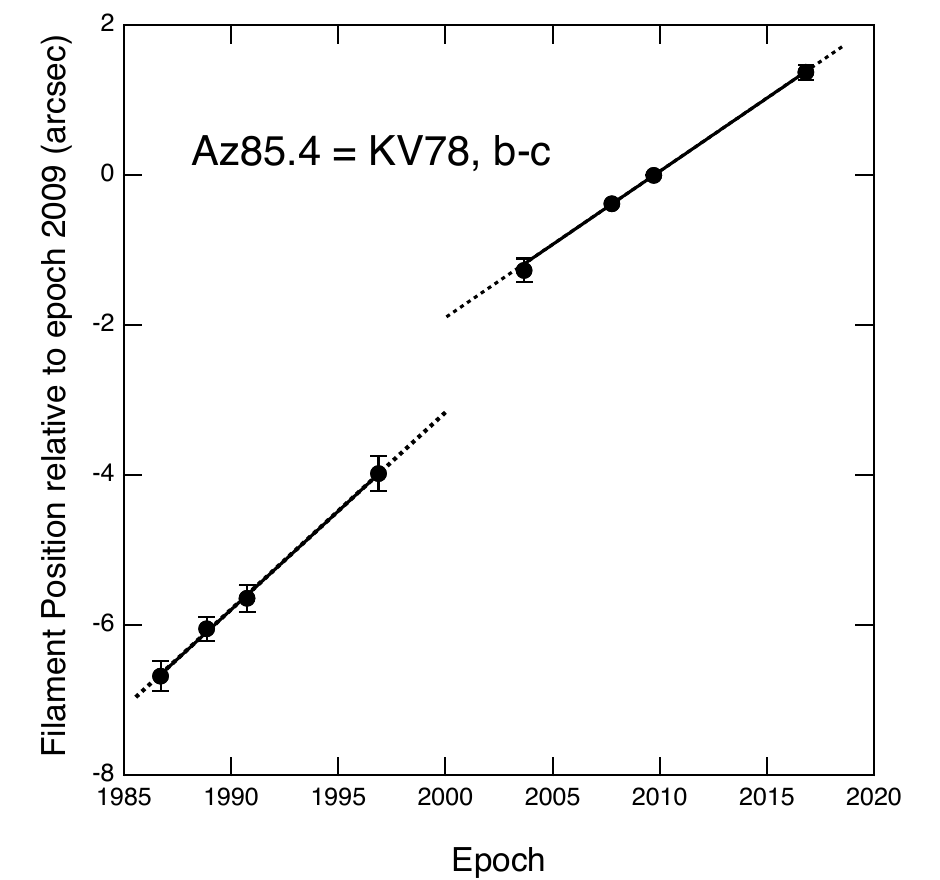}
\caption{\textit{Left}: An example of a filament that has been decelerating.  A linear fit was not acceptable, but the quadratic curve  shown here is much better. \textit{Right}: A filament whose motion has clearly been discontinuous. Both are features noted by KV78.
\label{plot_decel}}
\end{figure}

While we primarily selected filaments  with no obvious evolution for measurement,  we did measure some of these variable filaments for comparison with the results of KV78.
%when we did the fits to get the proper motions, we found that a few of them turned out to have nonlinear motions---acceleration/deceleration or discontinuities over our 30-year baseline.  
We  have included  these in Table \ref{tbl-results}, but give proper motion values only for ones where the motion has been linear over the last four epochs.  We show two examples in Fig.\ \ref{plot_decel}.  Both these examples appear in Fig.\ \ref{E_rim} and can be seen in KV78 Fig.2: Az79.9 corresponds approximately to KV78 feature d, and Az85.4 to their feature b-c.  In Table\ \ref{kv78_comp} we compare our measurements with those of KV78 for segments where we have measurements in common.  
\added{We can say little about the comparative directions  for motions measured by KV78 and ourselves.  KV78 give only the magnitudes for the filament motions they have measured.  We show the motion directions in Fig.~\ref{extrapolation}, where we estimate that these directions have uncertainty about $\pm 2\degr$.}

\begin{comment}

\begin{figure}[b!]
\epsscale{0.50}
\plotone{ movie_north.jpeg}
%\plottwo{E_bright_movie.pdf}{E_faint_movie.pdf}
\caption{Similar to Fig.\  \ref{E_rim}, except for fainter filaments along the northern rim.  Only six epochs are shown since the earliest epochs did not cover the full field.  The field of each individual image is $150\arcsec \times 40\arcsec$,  oriented N up, E left.
\label{N_rim}}
\epsscale{1.0}
\end{figure}
\nopagebreak[4]

\end{comment}

%For accelerating filaments or ones with discontinuous motion, we report measurements only where the motion appears uniform over at least the last four epochs: 2003-2016. 

\begin{deluxetable}{cccccc}[b!]
%\rotate
\tabletypesize{\small}
\tablewidth{0pt}
\tablecaption{Comparison of KC78 results  with the present ones \label{kv78_comp}}
\tablehead{
		\multicolumn{3}{c}{Kamper \& van den Bergh (1978)}  & & \multicolumn{2}{c}{Present paper}\\
		%\multicolumn{3}{c}{{Kamper and van den Bergh (1978)}  & \multicolumn{2}{c}{Present paper}\\
		%\cline(1-3} \cline{4-5}
		\colhead{Designation}  & 
		\colhead{Azimuth\tablenotemark{a}}  &
		\colhead{ $ \phn\phn \mu\;(\arcsec\,\peryr$)} &     &
		\colhead{Azimuth\tablenotemark{a}}  &
		\colhead{ $ \mu\;(\arcsec\,\peryr)$}    
		}
%		\decimals
\startdata
\vspace{-.2cm} a & 88.5 & $0.18 \pm 0.09 $ & & &  \\  
\vspace{-.2cm} & & & & 90.9 & decelerates/evolves   \\
\vspace{-.2cm} b & 87.5 & $0.15 \pm 0.05 $ & & & \\ 
\vspace{-.2cm}  & & & & 85.4 & decelerates/evolves   \\ 
\vspace{-.2cm}c & 79.8 & $0.26 \pm 0.09 $ & & & \\
\vspace{-.2cm} & & & & 79.9 & decelerates/evolves   \\ \vspace{.5cm}
\vspace{-.2cm}d & 78.2 & $0.20 \pm 0.08 $ & &  &  \\
\vspace{-.2cm} e & 76.5 & $0.22 \pm 0.01 $ & & & \\
\vspace{-.2cm} & & & & 77.5 & \phn $0.204 \pm 0.011$\tablenotemark{b}  \\ \vspace{.5cm}
f & 75.7 & $0.21 \pm 0.02 $ &  & & \\ \vspace{.5cm}
g & 70.5 & $0.20 \pm 0.01 $ &     \phn\phn  & 74.6 & \phn $ 0.157 \pm 0.013$\tablenotemark{b}  \\ \vspace{.5cm}
h & 15.5 & $0.28 \pm 0.05 $ & & 13.3 & $ 0.208 \pm 0.007 $ \\ 
\vspace{-.2cm}i &342.2 & $0.22 \pm 0.05 $ & & & \\
\vspace{-.2cm} & & & & 339.4 & $ 0.234 \pm 0.008 $   \\ \vspace{.5cm}
j &341.2 & $0.20 \pm 0.02 $ & &  &\\
\vspace{-.2cm}k &340.4 & $0.28 \pm 0.03 $  & & & \\
\vspace{-.2cm} & & & & 336.9 & $ 0.221 \pm 0.008 $   \\ \vspace{.5cm}
l &339.5 & $0.25 \pm 0.05 $ & &  & \\
\vspace{-.2cm}m &336.3 & $0.26 \pm 0.03 $ &  &  &\\
\vspace{-.2cm} & &  & & 334.4 & $ 0.218 \pm 0.008 $   \\ \vspace{.2cm}
n &334.4 & $0.18 \pm 0.06 $ &  &  &  \\
\enddata
\tablenotetext{a}{Our azimuths differ from those of KV78 because we chose significantly different center positions.}
\tablenotetext{b}{Also decelerates; proper motion was measured for epochs 1990-2003 only; see Sec.~\ref{distance}.}
\end{deluxetable}

\nopagebreak[4]

%\todo[inline]{Table 3 still needs a bit of editing.}

%==============================================================

\begin{figure}[h!]
\plotone{ 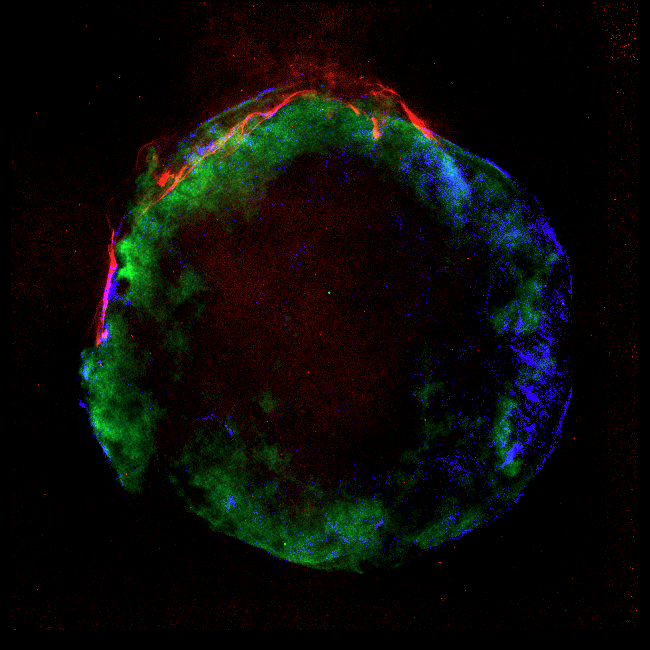}
\caption{Multi-wavelength image of Tycho's SNR: red is our 2009 continuum-subtracted H$\alpha$ image, green is a 1375 MHz radio image from the VLA in 1994-1995, which has been expanded by 14.7 years at 0.1126\%/yr, the mean rate of expansion reported by \citet{reynoso97}, and \added{blue is a deep, hard-band (3-6 keV)  X-ray  image from $Chandra$ in 2009 (Seq. 501025, J. Hughes, PI).}\label{fig-multi_image}}
%The centers used by these X-ray and radio studies (for determining expansion indices) are marked. 
%\todo[inline]{Joe has said blue is X-ray, and green is radio.  That's what I think we SHOULD have, but this figure is clearly as described in the caption.  I also want to get the later X-ray and radio images from Brian Williams.}\label{fig-multi_image}}
\end{figure}

\begin{figure}[h!]
\plotone{ 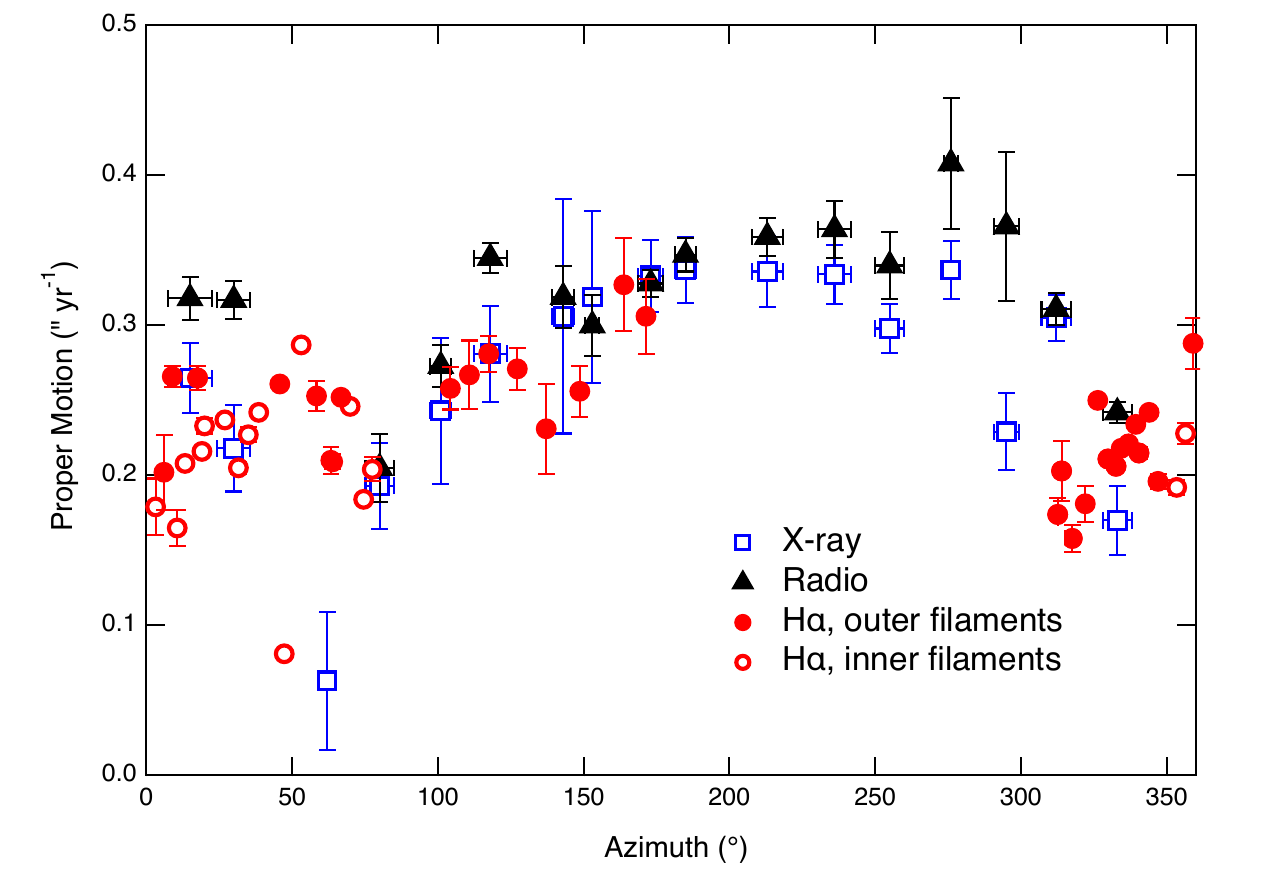}
\caption{Proper motion measurements in X-rays, radio (both from \citet{williams16}, and \ha\ (this work). For the \ha\ measurements, we distinguish between ones for filaments lying very near to the outer rim of the SNR shell, as delineated in X-ray/radio images, and ones located more to the interior.  \label{multilam_pm}}
\end{figure}

\begin{figure}[b!]
\epsscale{1.15}
\plotone{ 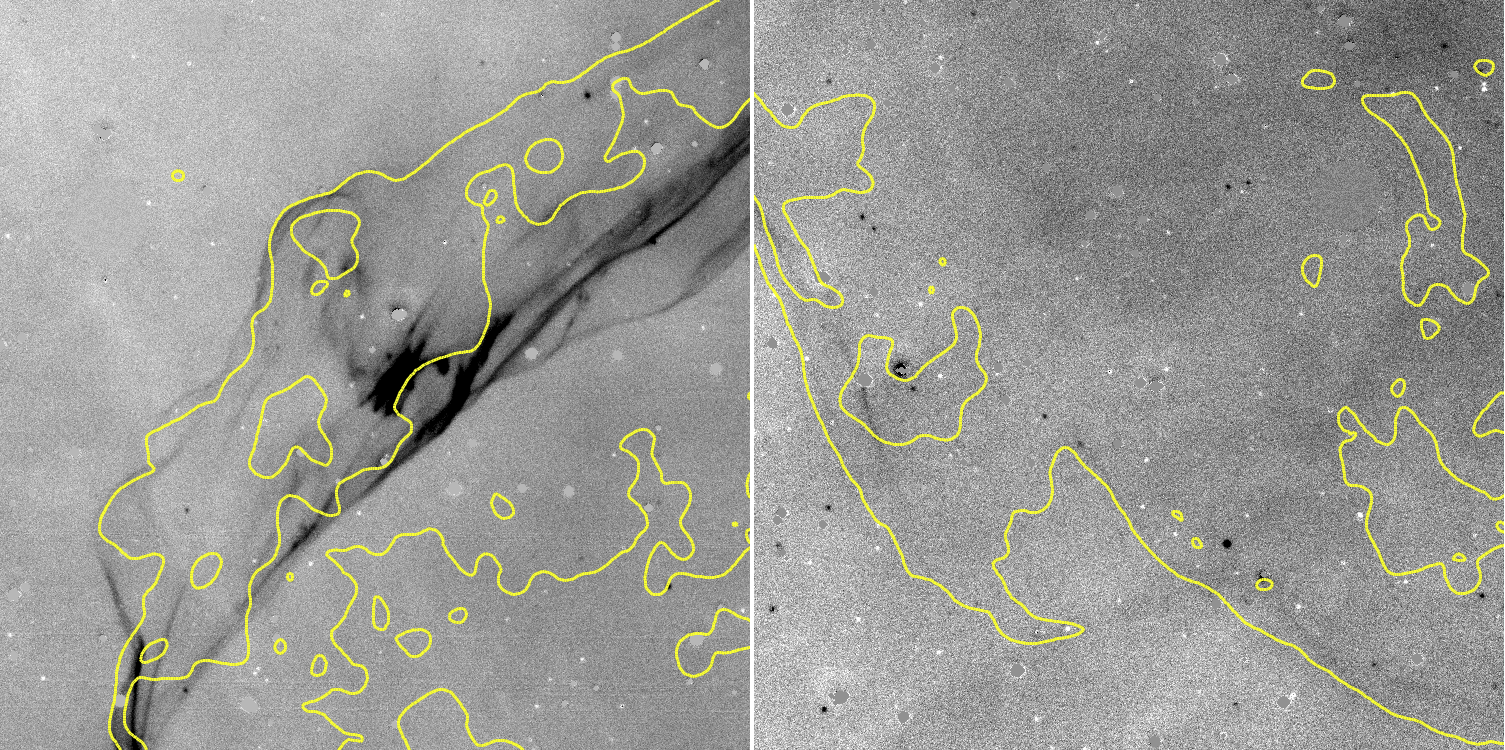}
\caption{\ha\ image of Tycho (epoch 2009) with low-level contours from the 2007 hard X-ray image overlaid.  {\em Left:} NE region,  
where the bright \ha\ filaments lie a few arcsec within the X-ray rim, but there are fainter filaments near-coincident with the X-ray rim.  {\em Right:} SE region, where the extremely faint \ha\ emission coincides well with the X-ray rim.  Both fields are 2.5\arcmin\ square, oriented N-up, E-left. \label{fig-contours}}

%The centers used by these X-ray and radio studies (for determining expansion indices) are marked. 
%\todo[inline]{Joe has said blue is X-ray, and green is radio.  That's what I think we SHOULD have, but this figure is clearly as described in the caption.  I also want to get the later X-ray and radio images from Brian Williams.}\label{fig-multi_image}}
\end{figure}

\section{Multi-Wavelength Comparison \label{multi}}

%\subsection{}
%==============================================================
As noted in Sec.\ \ref{intro}, Tycho has long been known to be a source of radio and X-ray emission, and its shell is notably circular in both these bands.  Tycho's radio and its hard  X-ray emission both arise from the same fundamental physical process: synchrotron radiation from highly relativistic electrons that have been accelerated in the turbulent magnetic field behind the primary SN shock.\footnote{While the X-ray emission from most SNRs is dominated by thermal emission lines, \citet{hwang02} have shown that the 4-6 keV X-ray  emission from Tycho is almost entirely  non-thermal.} However, the post-shock profiles are quite different in the two bands.  The hard X-ray emission peaks sharply in a thin outer shell only about 4\arcsec\ thick immediately behind the shock, while the radio shell is much thicker and more diffuse \citep{hwang02}.  In Fig.\ \ref{fig-multi_image} we show an image of Tycho color-coded to illustrate the emission in the different energy bands.

%\todo[inline]{There have been 3 recent papers on X-ray expansion since W16.  I should mention some or all,}

There have also been several radio \citep{reynoso97} and X-ray \citep{hughes00,katsuda10} proper-motion studies of Tycho.  The most recent expansion measurement at the rim of Tycho in both bands has been done by \citet[][henceforth W16]{williams16}, who also summarize  earlier studies.  Fig.\ \ref{multilam_pm} reproduces Fig.\ 4 from W16, and we have added our optical measurements (Table \ref{tbl-results}).  The X-ray and optical proper motions are  quite well correlated over most of the azimuthal range where we see optical filaments.  While also roughly correlated, the radio motions are somewhat faster than those seen in the X-ray or optical.  More recently, \citet{tanaka21} have carried out  another X-ray expansion study of Tycho, using four epochs of {\em Chandra} data, and they find that the expansion of the hard X-ray rim has recently decelerated, at least in the east through southwest directions (azimuths $\sim 80\degr-260\degr$).  This is  consistent with our findings for the optical filaments, where we also observed deceleration for the bright eastern filaments, though we cannot say whether their deceleration has been as sudden as Tanaka et al. find in X-rays.  

At azimuths $\gtrsim 100\degr$ the \ha\ filaments are so faint that we detected them only in our more recent  observations (if at all), so we would not have been able to observe possible deceleration.  Tanaka et al.\ interpret the recent deceleration seen in X-rays as being due to the primary SN shock having recently hit the dense wall of a pre-existing cavity, which may have been formed by strong winds from a singly-degenerate progenitor. %A recent X-ray spectroscopic study from {\em XMM-Newton} by \citet{uchida24} is  consistent with this basic picture.
But the optical data can  equally well be interpreted as simply higher pre-shock density to the east---perhaps due to the presence of a dense cloud.
Our study, like the proper motion measurements in radio and X-rays, shows that Tycho is expanding fastest in the southeast, south, and southwest (azimuth $\sim$ 110\degr - 300\degr), where we see only very faint or no  optical emission. Slower expansion at azimuths with prominent optical filaments suggests that Tycho's expansion has encountered higher density regions in the east and north. 
%(azimuths $\sim$ 0\degr - 90\degr and 330\degr - 360\degr). It is important to compare our proper motion results for optical filaments coinciding with the X-ray/radio rim. We first created a multi-wavelength image as shown in Figure~\ref{fig-multi_image} using our 2009 epoch, a broad-band X-ray image taken by Chandra in 2007 (see K10), and a 1994-1995 radio image from the VLA expanded by 14.7 years at 0.1126 \%/yr,the mean rate of expansion reported by R97. (Also, R97 shows a radio-optical image of Tycho using our 1990 epoch.) The main areas where some optical filaments coincide with the X-ray/radio rim is in the east and northwest. Most of the north filaments are interior to the rim, with the exception of one faint filament (filament 9). In the north, there is very diffuse H$\alpha$ emission along the rim and in the pre-shock environment. 

%The most recent X-ray and radio expansion measurements are given in W16, who used time baselines of 12-15 yr and 30 yr, respectively. 

In Fig.~\ref{fig-contours} we show regions of our 2009 \ha\ image with low-level contours of the hard X-ray image (also from 2009) overlaid.  In the northeast (left panel), the bright optical filaments lie 30\arcsec - 40\arcsec\  behind the outer X-ray rim.  However, there are much fainter optical filaments near-coincident with the X-ray rim.  We interpret this as likely due to projection effects from the 3-dimensional structure of Tycho; the X-ray and faint optical filaments are formed from interaction of the SN shock with low-density gas at the projected rim of the 3-D shell, while the brighter \ha\ filaments represent regions where the shock is moving near transverse to the line of sight through denser regions of the ISM.  In the southeast and south (right panel), the optical emission is extremely faint and less sharp.  It is concentrated in a band about 4\arcsec - 10\arcsec\ behind the X-ray rim, which itself is relatively faint.  This probably is due to a very low pre-shock density.  This is consistent with the higher proper motions  observed in both the X-ray and optical filaments at azimuths $\sim150\degr - 240\degr$ (Fig.~\ref{multilam_pm}).

%\todo[inline]{The following is from Joe. I think the most important thing for us is to do a plot like W16 Fig. 4, adding the optical PM's.  Perhaps we might distinguish between filaments at the radio/X-ray limb, vs ones further inside.  In any case I plan to experiment with this.}

%W16 measured proper motions for 17 different sections throughout the rim. Section 6 is representative of our filament 1, section 4 of our filament 4, and section 17 of our filament 28. W16 found a discrepancy in the X-ray and radio proper motions for their section 6. Our filament 1 measurement is consistent with their X-ray measurement, but their radio measurement is less uncertain than their X-ray measurement. X-ray, radio, and optical proper motions for section 4 all agree. W16 found a significant discrepancy in the X-ray and radio proper motions for their section 17.  Our filament 28 measurement is consistent with their radio measurement, which has a much smaller uncertainty than their X-ray measurement. 

%==============================================================

\section{Distance to Tycho \label{distance}}
If one can measure the shock velocity $v_s$ in the \ha-emitting neutral H, together with the proper motion for SNR filaments moving  in the plane of the sky at the same location, one can obtain the distance  to the SNR purely geometrically, $\mu \equiv \dot{\theta} = v_s/d$. This was first pointed out by \citet{chevalier80}, who measured the width of the broad component of the Balmer lines from  knot g of KV78 and the broad/narrow intensity ratio. %, to determine the shock velocity. 
The velocity width of the broad component, $v_B$, gives a direct measure of the post-shock proton temperature, which is in turn related to $v_s$. 
%equal to the post-shock speed, $v_0 \approx \frac{3}{4} \,v_s $, where $v_s$ is the shock velocity.  
\added{There are a number of significant effects that affect the relation between  shock velocity and Balmer width, including the degree and timescale at which post-shock electrons and protons equilibrate in temperature,  the loss of energy into cosmic rays, and the extent of a possible shock precursor \citep{heng10}.}  These were estimated by Chevalier et al., and later  in increased detail by several authors \citep{kirshner87, smith91, ghavamian01, lee07, lee10, morlino13}.  Several of the latter authors obtained new spectra of Tycho, but \textit{all of these were taken from nearly the same spatial location,} KV78's knot g, and they have all used the KV78 value for the proper motion.  This is no doubt because this is the brightest knot in Tycho---at least it was in 1978 and for some years thereafter.

As we indicate in Fig.~\ref{E_rim} and Table \ref{tbl-results},  knot g and its surrounding complex have evolved in morphology and brightness in recent years, and its motion has $not$ been uniform.  Hence it is unlikely that its proper motion at the time of the 1980-2013 spectroscopic observations referenced above was the same as that measured by KV78, leading to   distance estimates that are likely erroneous.  The filaments we have studied embrace a wide variety of physical conditions.  Spectra from several of these filaments should enable multiple checks on the models for obtaining the shock velocity, leading to multiple consistency checks and ultimately a more reliable distance determination to Tycho.   %There might also be the possibility of exploring  its 3-D kinematic structure.  %For the present, however, we are limited to suggesting this pathway to future researchers.   

For the present, we can use the spectroscopy of \citet{ghavamian01} to refine the distance measurement somewhat.  Ghavamian et al.\  obtained their long-slit  at the brightest part of KV78 knot g (which they denote as g1) in 1997.  We are able to reproduce the slit position shown in  their Fig.\ 3 almost exactly on our 1996-epoch image, and their g1  is virtually identical to our region Az74.6.  Ghavamian et al.\ find that for knot g1 the broad component width is $1765 \pm 110 \kms$, and that the broad-to-narrow intensity ratio is $I_b/I_n = 0.67 \pm 0.1$. They then calculate that $v_s = 2110  \pm 170  \kms$, taking into account both measurement and model uncertainties.  We find that knot g has decelerated substantially over our 30-year baseline, but  that over the epoch sequence 1990-1996-2003 (which brackets Ghavamian et al.'s spectroscopy) its motion was virtually constant:  $\mu = 0.157 \pm 0.013 \arcsec\, \peryr$.  Combining these values for $v_s$ and $\mu$ gives a distance  of $2.83 \pm 0.33 $ kpc.

Past estimates of Tycho's distance range widely, from 2 kpc to 5 kpc. 
As noted above, \citet{ chevalier80}, \citet{kirshner87}, and \citet{smith91}  have  all previously estimated the distance to Tycho from their measurements of broad Balmer profiles and the proper motion of knot g by \cite{kamper78}, so naturally they obtained similar results: $2.3 \pm 0.5$ kpc, 2.0 - 2.8 kpc,  and 1.5 - 3.1 kpc, respectively.  Distance estimates by  other methods include H\thinspace I absorption, X-ray kinematics, deduction based on the light echo together with the brightness of SNIa and reddening, and theoretical estimates based on the Sedov model or on X-ray hydrodynamics.  
%These range from 2 to 5 kpc.  
\citet{devaucouleurs85} reviewed distance estimates (and upper/lower limits) available at that time, and he summarizes by concluding that $d_{\text {Tycho}} = 3.2 \pm 0.3 $ kpc. \citet{hayato10}  have a nice graphical summary of distance measurements before 2010.  \added{More recently, \citet{williams13} found that a distance $d\sim 3.5$ kpc was most consistent with their IR observations of Tycho and standard models for dust grains.  \citet{slane14} measured the positions and expansion speeds of the forward shock and reverse shock, and find that Tycho is  consistent with their hydrodynamic model for a distance $d \sim 3.2$ pc. A number of recent studies have measured the 3D motions of ejecta clumps in Tycho, and have combined these with  the assumption of a near spherical ejecta distribution and/or hydrodynamic models to obtain a distance estimate:  \citep[][$d\sim 4.0$ kpc]{sato17},  %used {\em Chandra} ACIS and [\em Suzaku} data  
 \citep[][$d \sim 3.5$ kpc] {williams17} and \citep[][$d \sim 3.5$ kpc] {millard22}. %used the transmission gratings on {\em Chandra} to obtain a 3D map of the ejecta distribution, and assuming a roughly spherical distribution, they estimate a distance of 3.5 pc.

Our proper motion measurements for  a large number of filaments, encompassing a wide variety of physical conditions, offer the promise of a significant improvement in the distance measurement to Tycho.  If future observers obtain spectra that give Balmer profiles with reasonable precision, then it should be possible to calculate the  shock velocities for a variety of filaments, and thus reduce the model uncertainties that have limited the precision of geometric measurements, including our own.}

\section{Summary and Conclusions \label{summary}}
%==============================================================

(1) We have obtained CCD images of the Tycho SNR  at eight epochs extending back to the earliest common use of CCDs in astronomy: from 1986 through 2016.  These reveal a far more extensive network of \ha-emitting filaments than shown in previous photographic images.  The later epoch (more sensitive) images also show very faint, diffuse emission surrounding  the southern and southeastern rims of the SNR shell, azimuths $\sim 90\degr-180\degr $. These align closely with the radio/X-ray rim in the same locations.  In all, we now see optical emission around almost two-thirds of the circumference of the Tycho shell.

(2) The majority of the filaments have remained reasonably stable in structure, and have moved uniformly over our 30-year baseline.  However, our detailed images show that several of the brighter filaments have evolved noticeably in morphology and/or brightness over this period.  

(3) We have obtained spatial profiles transverse to many of the filaments and have measured the shifts from epoch to epoch.   Linear fits to the shifts as a function of time have enabled us to measure proper motions for 46 filaments, typically to a precision of 1\% -- 10\%.

(4) Our measurements agree  reasonably well with those from the previous photographic measurements by \citet{kamper78}, though we find that several of the brighter filaments in their study have changed significantly in morphology and/or decelerated.

(5) We find that the optical and hard X-ray expansion  are generally well correlated as a function of azimuth, especially  for those optical filaments closest to the shell rim.   The radio expansion is less well correlated with that measured either optically or 
in X-rays.

(6) Several of the optical filaments, especially ones in the bright  complex along the east rim of Tycho, have decelerated over the past thirty years.  This is consistent with the deceleration  of the near-coincident X-ray emission at the rim of Tycho to the east.   This could be due to the primary shock  hitting the wall of a pre-existing cavity formed by winds from a singly degenerate SN progenitor \citep{tanaka21}.  But a local ISM enhancement  could instead be responsible for the decelerating optical filaments.  \added{In the south and southeast, the optical filaments are very faint and are moving fastest, which is also consistent with the faint hard X-ray emission at the rim in this region.}

(7)  Most optical filaments at the shell rim have expansion indices, the ratio of current expansion rate divided by the average rate over the remnant's history,  $m = \mu/(r/t)$,  relatively consistent with the Sedov value of 0.40.  Interior filaments generally have smaller values, as expected. Interior regions may have been decelerated, and/or they could be moving non-tangentially to the line of sight.

(8) Measurement of the angular expansion of the shell rim and the shock velocity for the shocks driving that expansion would enable a purely geometric determination of the distance to Tycho.  The width of broad Balmer-line profiles, resulting from charge exchange between low-temperature pre-shock neutral atoms and hot post-shock protons, can give a measure of the shock velocity. Applying  this method to KV78 knot g,  taking $ v_s $ from \citet{ghavamian01} and combining that with our {\em contemporaneous} proper motion measurement,  we obtain an estimate of $d_{\text {Tycho}}  =  2.83 \pm 0.30 $ kpc.   
However, the models describing this process  have multiple uncertainties that limit  the precision of $v_s$, and hence the distance determination.  We hope that future observers will obtain spectra of additional filaments, which can be combined with our proper motions and with model refinements to refine the distance measurement in future.

\bigskip

Our complete set of aligned images at all eight epochs, both original (with the stars) and continuum-subtracted, are accessible at the Middlebury Optical ATlas of SNRs (MOATS), https://sites.middlebury.edu/snratlas/tycho-snr/ . \added{The same data may be accessed on the Harvard AAS Dataverse site https://doi.org/10.7910/DVN/QSGIMF.}  At the MOATS site is also a 2.5 min video showing our sequence of Tycho images in animated form.

\bigskip

\begin{acknowledgments}
%==============================================================
We are extraordinarily grateful to Rob Fesen for sharing his 1988 and 2003 images of Tycho.  We also thank Ben Williams and Kristoffer Eriksen (1996) and Matt Vaughn (2007) for their assistance with  observing runs  and for aspects of the data processing. Parviz Ghavamian has helpfully explicated details from his spectra of Tycho filaments.  We thank the anonymous referee for an extremely careful review of the manuscript, which resulted in numerous improvements to this paper.  PFW  acknowledges financial support from the NSF through grant AST-098566, as well as through additional grants in earlier years.  WPB acknowledges support from the JHU Center for Astrophysical Sciences.  Finally, the Middlebury College faculty research fund has partially underwritten the costs of publishing this research. 

\facilities {Gemini:North (GMOS), WIYN  3,5m, KPNO Mayall 4m, KPNO 2.1m, MDM Hiltner 2.4m}

\software {SAOimage ds9, IRAF, Igor Pro}

Supported by NSF grant AST-098566.
\end{acknowledgments}
%- Made use of image display tool SAOImage DS9 developed by the Smithsonian Astrophysical Observatory
%==============================================================

\bibliographystyle{aasjournal}

\bibliography{bibmaster}

\end{document}